# The ERA2 facility: towards application of a fiber-based astronomical spectrograph for imaging spectroscopy in life sciences

Martin M. Roth[a,b], Karl Zenichowski[a], Nicolae Tarcea[c], Jürgen Popp[c,d], Silvia Adelhelm[a], Marvin Stolz[a], Andreas Kelz[a], Christer Sandin[a], Svend-Marian Bauer[a], Thomas Fechner[a], Thomas Jahn[a], Emil Popow[a], Bernhard Roth[e], Paul Singh[a], Mudit Srivastava[a], Dieter Wolter[a]

[a] Leibniz-Institut für Astrophysik Potsdam (AIP), An der Sternwarte 16, 14482 Potsdam, Germany; [b] Universität Potsdam, Institut für Physik und Astronomie, Karl-Liebknecht-Str. 25/26, 14476 Potsdam, Germany; [c] Institut für Physikalische Chemie, Universität Jena, Helmholtzweg 4, 07743 Jena, Germany; [d] Institut für photonische Technologien, Albert-Einstein-Str. 9, 07745 Jena, Germany; [e] Hannoversches Zentrum für optische Technologien (HOT), Nienburger Str. 17, 30167 Hannover, Germany

## ABSTRACT

Astronomical instrumentation is most of the time faced with challenging requirements in terms of sensitivity, stability, complexity, etc., and therefore leads to high performance developments that at first sight appear to be suitable only for the specific design application at the telescope. However, their usefulness in other disciplines and for other applications is not excluded. The ERA2 facility is a lab demonstrator, based on a high-performance astronomical spectrograph, which is intended to explore the innovation potential of fiber-coupled multi-channel spectroscopy for spatially resolved spectroscopy in life science, material sciences, and other areas of research.

## 1. INTRODUCTION

Fundamental research nowadays faces expectations from industry, society, and politicians to not only address primary science goals, but also to be open for inter- and multidisciplinary collaboration, and to be ready to provide knowledge and technology transfer into other disciplines of research, as well as into the commercial sector. While it is not always clear whether for a given technology a transfer potential really exists, there are others where one must really suspect that this is indeed the case. However, in the latter instance the assumed transfer case is not readily validated. Not only does the exploration of technology transfer opportunities require substantial human resources, money, and time, but also is the potential recipient of a transfer activity not necessarily prepared to absorb and subsequently exploit the new technology. The MIT School of Engineering (MIT 2012) identifies four major obstacles that are thought to be responsible for the existence of the so-called innovation gap:

- Fear of risk
- Reduced funding
- Financial limitations on small business
- Disconnect between academia and marketplace.

Although its mission remains to be rooted in fundamental astrophysical research, the Leibniz-Institut für Astrophysik Potsdam (AIP) has made an effort to help overcoming the innovation gap for potentially worthwhile developments of its R&D portfolio by creating the *Leibniz Application Lab for fiber-optical Spectroscopy and Sensing*, which is one of three major activities of the innovation center *innoFSPEC Potsdam*, whose headquarter is located at the Babelsberg campus of AIP (Roth et al. 2008). Chiefly, the application lab is providing access to lab demonstrators in order to encourage potential external users to gain hands-on experience with the featured technologies, and to facilitate the assessment of any chances for transfer into other disciplines, or even for commercialization. A first realization of this vision is the implementation of a fiber-coupled spectrograph test bench called ERASMUS-F, whose primary purpose is to investigate the long-term behaviour of optical fibers and fiber bundles for multi-object spectroscopy with single fibers, or for deployable integral-field units (IFUs), respectively. Here we describe the lab demonstrator setup, and our very first experiments towards the validation of the transfer potential for multiplex Raman spectroscopy (MRS), laser-induced breakdown spectroscopy (LIBS), and other future applications.

# 2. TECHNOLOGY TRANSFER CONCEPT

In the following, some analogies between spatially resolved spectroscopy in astrophysics and in life science will be highlighted in order to demonstrate why there is a strong case for technology transfer from applications of multi-channel spectroscopy in astronomy to other disciplines.

## 2.1 Integral Field ("3D") Spectroscopy

Integral field spectroscopy, often also called 3D or IFU spectroscopy, is a still relatively new, however well-established observing technique in astronomy (Roth 2010). After a long period of introduction and prototyping, nowadays all major ground-based observatories do offer 3D spectrographs in the visible and near-infrared wavelength regions, with 2$^{nd}$ generation instruments on the horizon, e.g. at the ESO-VLT (Bacon et al. 2010), Keck Observatory (Morrissey et al. 2012), or in space (Prieto 2003).

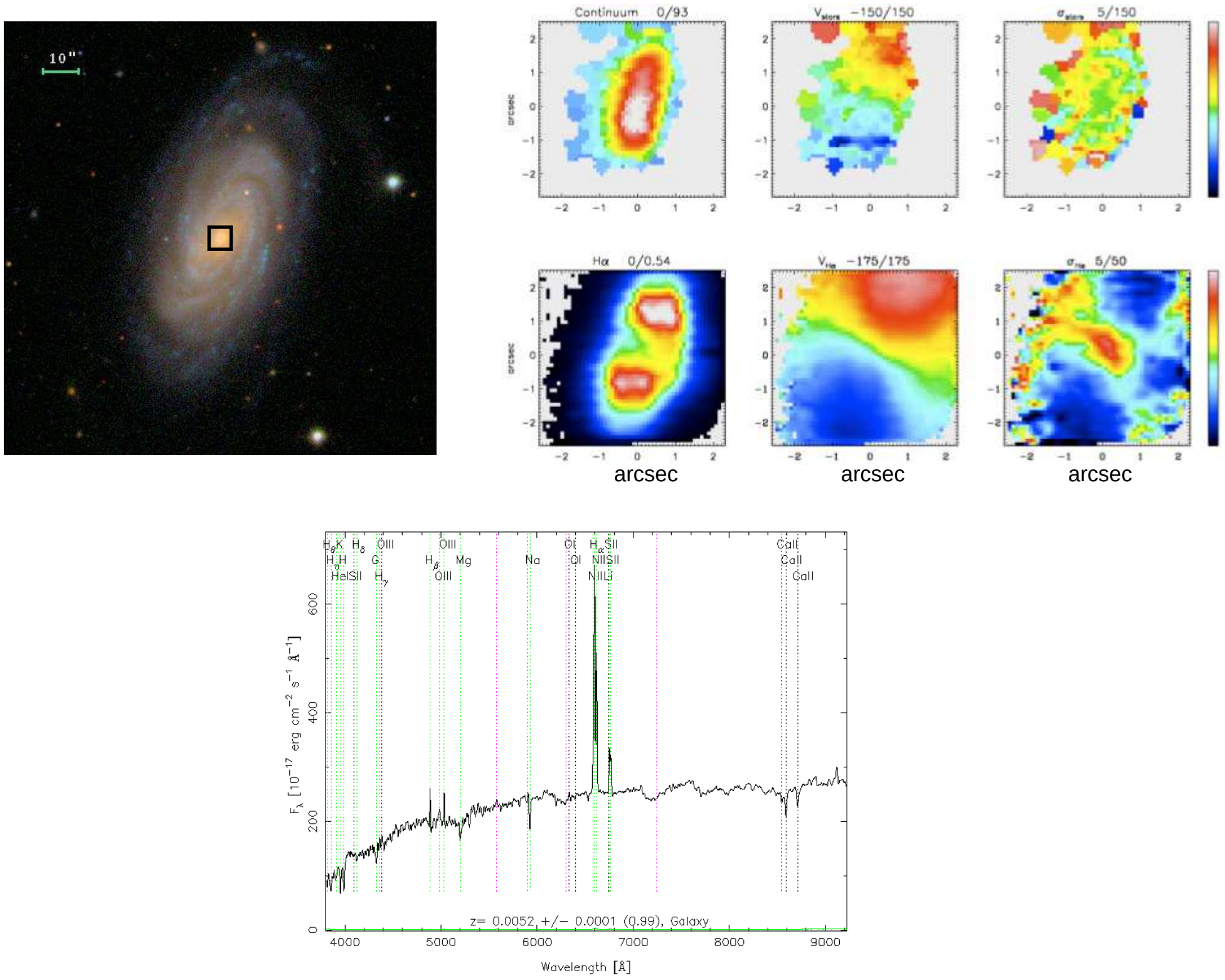

Figure 1: IMACS integral field spectroscopy of the NGC 5740 (from Westoby et al. 2012), spectrum from SDSS DR7

Fig.1 illustrates the typical data product of this technique, which happens to come as a data cube: the square in the direct image to the upper left indicates the field-of-view of the IMACS IFU, while the colour-coded frames to the right show maps in continuum light, the velocity field of the stellar absorption line component, and the corresponding velocity dispersion in the upper row, and maps of hydrogen Hα emission line intensity, the Hα velocity field, and again velocity dispersion, correspondingly. All of these maps have been derived from a datacube. A typical spectrum from the galaxy that could be taken from the datacube is also shown on the bottom of Fig. 1. Both the stellar continuum and absorption line component as well as the gaseous emission line component of this galaxy are clearly visible.

## 2.2 Imaging Raman Spectroscopy

While integral field spectroscopy as such is not established yet in biology, medicine, or other life science areas at the same level as in astronomy, there is a clear demand for spatially resolved spectroscopy in order to map organic tissue and to derive specific information about tissue properties, e.g. pathologically modified cell structures, on a pixel-by-pixel basis. In a review article on this subject, Krafft et al. (2009) describe a method based on fiber-coupled Raman spectroscopy which is capable of achieving this goal, however as yet merely using a scanning technique that requires a tedious procedure of point-wise sampling the object under study. Fig. 2 presents an example of this technique, namely a microscopic image of a lung fibroplast cell (left), a Raman image (right) obtained from raster sampling spectroscopy, and subsequent cluster analysis of spectra as shown in the bottom frame. The colour code of the Raman image indicates spectroscopic identification of the nucleus (red), the cytoplasm (cyan), and lipid vesicles (green).

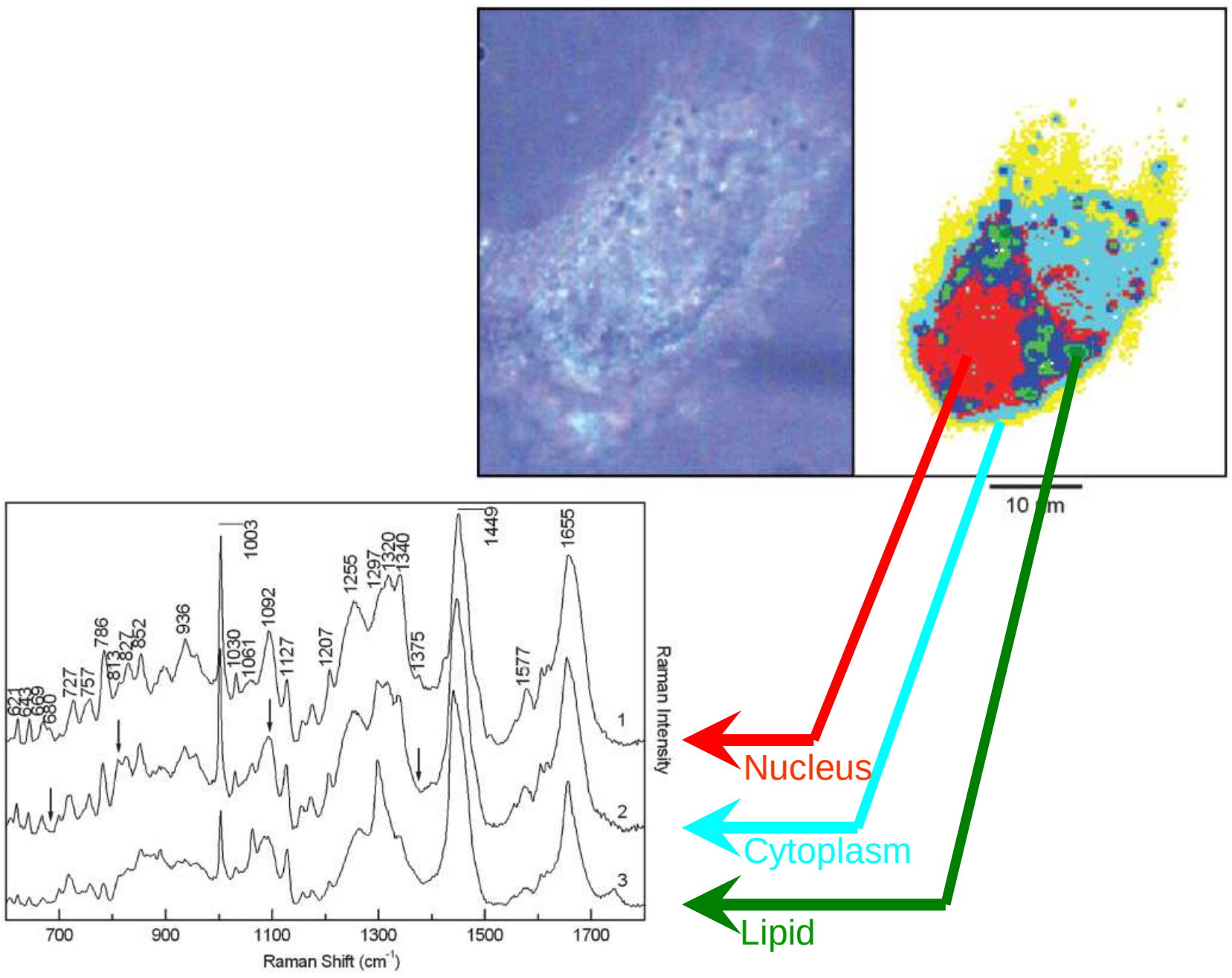

Figure 2: Imaging Raman spectroscopy of a lung fibroplast cell (from Krafft et al. 2009)

Comparing Fig.1 and Fig.2, the similarity is striking: images, maps, and individual spectra of a galaxy versus the same representations for an organic cell. Obviously the analysis of the spectral signature is completely different, however, the format of data entities practically identical. The major difference between the two datasets is the way how they were created in the measurement process: data for the galaxy in Fig.1 were created from integral field spectroscopy, i.e. basically from an exposure in one shot, whereas data for the cell in Fig.2 resulted from a scanning series of individual spectra, taken one by one for each spatial element. The logical conclusion from this observation is the question: is it possible (and desirable) to transfer the fast integral field approach in astronomy to life science, and thus to replace the time consuming scanning process by a single data acquisition? The lab demonstrator and experiments explained in the following sections are intended to exactly address this question and to validate the notion that the effort might be justified.

### 2.3 Technology transfer demonstrator at the Leibniz Application Lab at AIP

As part of its technology transfer activities, AIP offers access to selected experiments and lab demonstrators to external parties who have an interest to explore astronomical technologies for other applications. This activity is known as the *Leibniz Application Lab for fiber-optical Spectroscopy and Sensing* (WGL 2012), which is embedded in similar activities of other institutes under the umbrella of the Leibniz-Gemeinschaft in Germany, with a common branding and corporate identity. Currently the application lab at AIP has a special focus on fiber-optical multi-channel spectroscopy, which is also one of the major areas of R&D for astrophysical instrumentation. It is tightly connected to an ongoing research activity at AIP, known as the *ERASMUS-F Spectroscopic Test Facility*, whose purpose is the systematic study of the long-term behavior and stability of fiber bundles for multi-object spectroscopy with deployable IFUs. Instrument concepts with this layout are gaining more and more attention in the scientific community, with new implementations being planned (e.g. MaNGA for the AS3 phase of SDSS), or already in operation (SAMI at the AAT, Bryant et al. 2012, this conference). Another concept proposed for the VLT, called FIREBALL, was presented by Kelz et al. 2010. The latter was based on a scalable modular layout with a modified MUSE spectrograph for fiber feed, which is also the technology used for the ERASMUS-F facility. The technology transfer demonstrator of the Leibniz Application Lab, for short dubbed ERA2, is as well making use of this hardware. Fig.3 shows the MUSE spectrograph mounted at AIP, originally for the purpose of acceptance testing for the 24 detector vessels of MUSE. However, as a matter of seamlessly swapping the collimator CCD test illumination unit for a fiber feed, the spectrograph could be used over a period of 2 days for tests as described below.

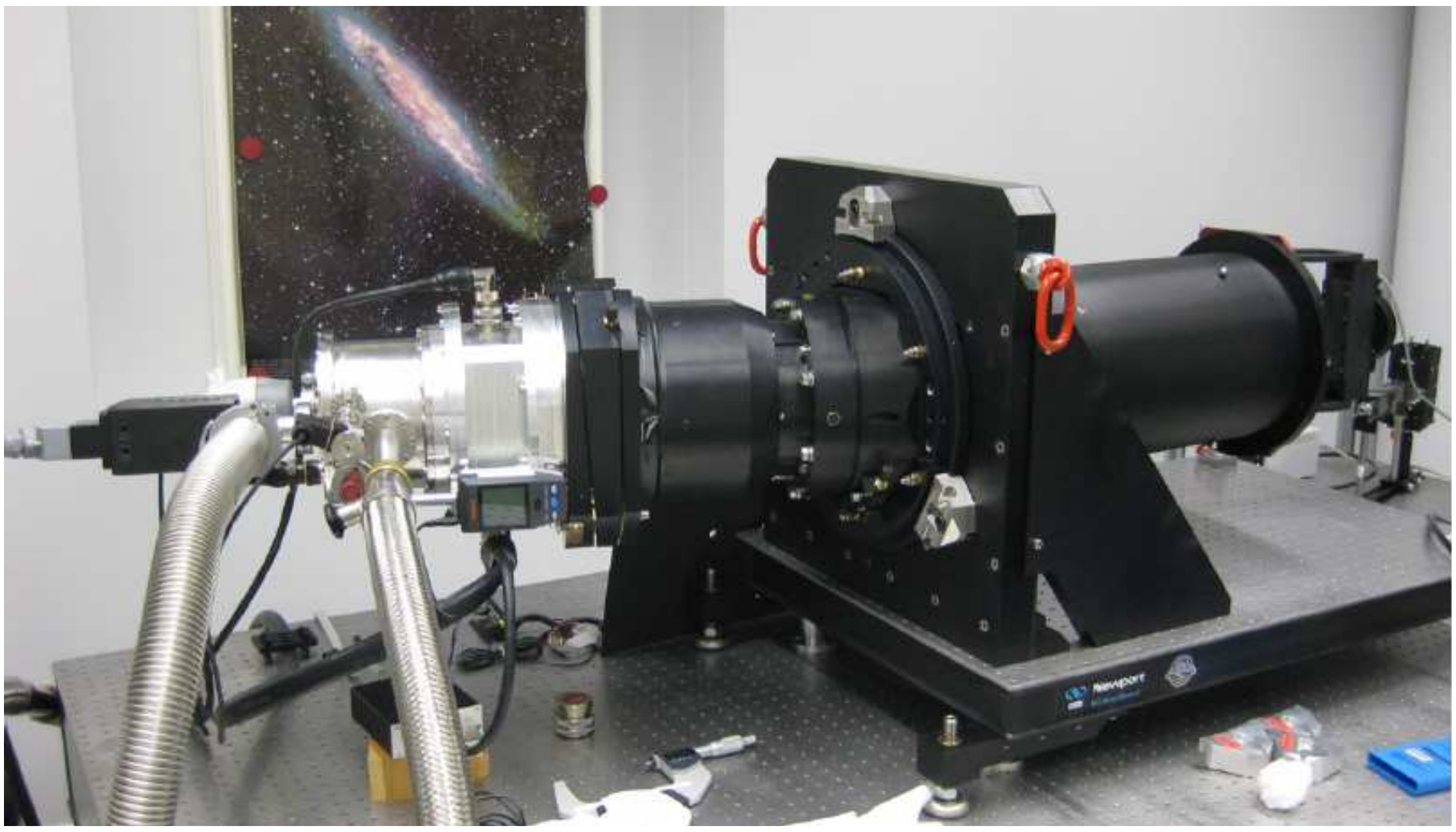

Figure 3: MUSE spectrograph in the *Leibniz Application Lab for fiber-optical Spectroscopy and Sensing* at AIP

The MUSE spectrograph is a high-throughput, high image quality optical system with a free spectral range covering one octave from 465nm – 930nm. It is intended to provide ultra-high stability for reliable calibrations and thus incorporates no moving parts. As the optical design is matched to the anamorphic MUSE fore-optics and an advanced image slicer system on spectrograph input, the pupil is elliptical rather than circular. The detector is an e2v 4K×4K CCD231 chip with an optimized graded index AR coating. A full description of the system and its performance when coupled to optical fibers is given in Kelz et al. 2010.

## 3. FIRST EXPERIMENTS WITH IMAGING RAMAN SPECTROSCOPY

A first series of tests was devised in order to validate the usefulness of the ERA2 facility for imaging Raman spectroscopy for life sciences and material sciences. To this end, the MUSE detector vessel test spectrograph was temporarily equipped with a fiber bundle and a pseudo-slit-collimator interface to replace the standard pinhole illumination unit employed for the MUSE detector acceptance tests. This setup was used for a test campaign on May 3 and 4, 2012. In what follows, the lab setup will be described, including the scope of the tests, and a summary of first results.

## 3.1 The ERA2 fiber-optical interface

In order to provide the ERASMUF-F test facility with a first set of fiber optics, a fiber bundle was designed with specifications derived directly from VIRUS bundles that were built and tested previously at AIP for use in the Hobby Eberly Dark Energy Experiment (Hill et al. 2010). This concerns mostly the fiber head design including a drilled hole mask to form the IFU matrix, the fiber bundle conduit to protect individual fibers and minimize the occurrence of strain and bending, the fiber slit in form of a v-groove steel substrate, and all fibers of the bundle glued to the substrate to form a linear array.

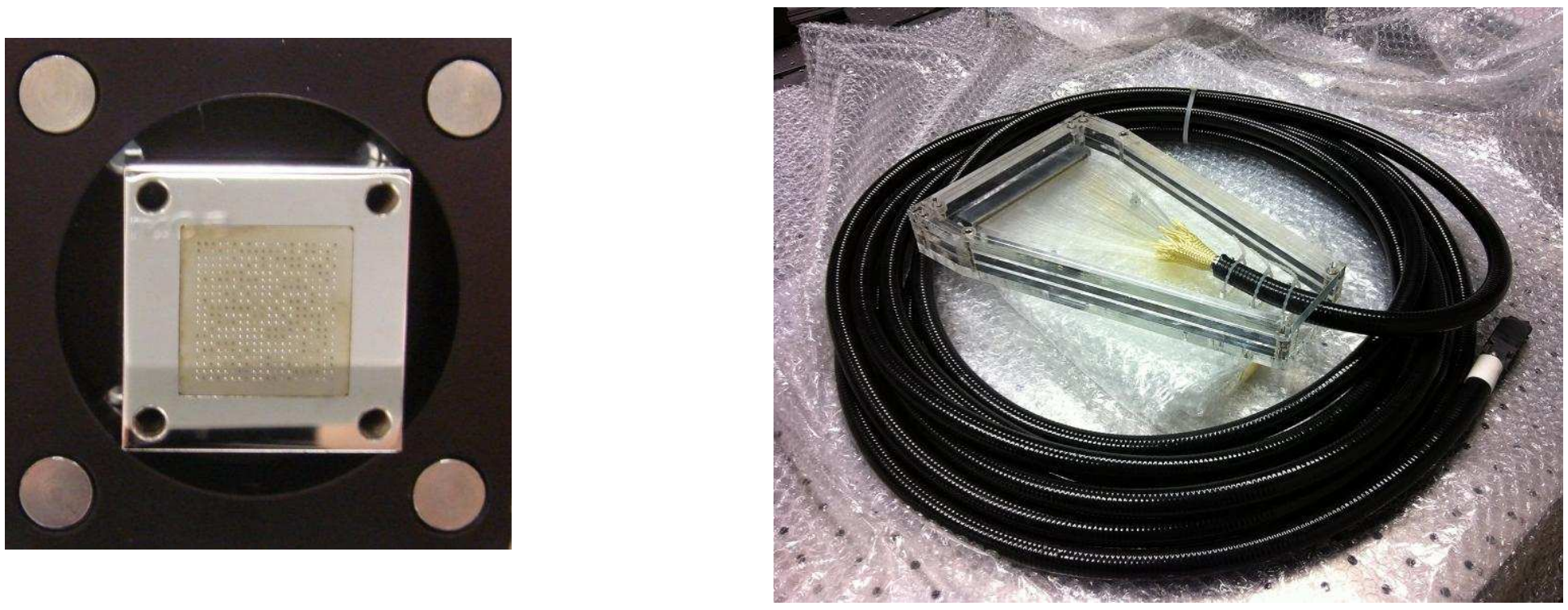

Figure 4: ERA2 fiber bundle (right), the partially back-illuminated fiber IFU head (top view) is shown in the left image

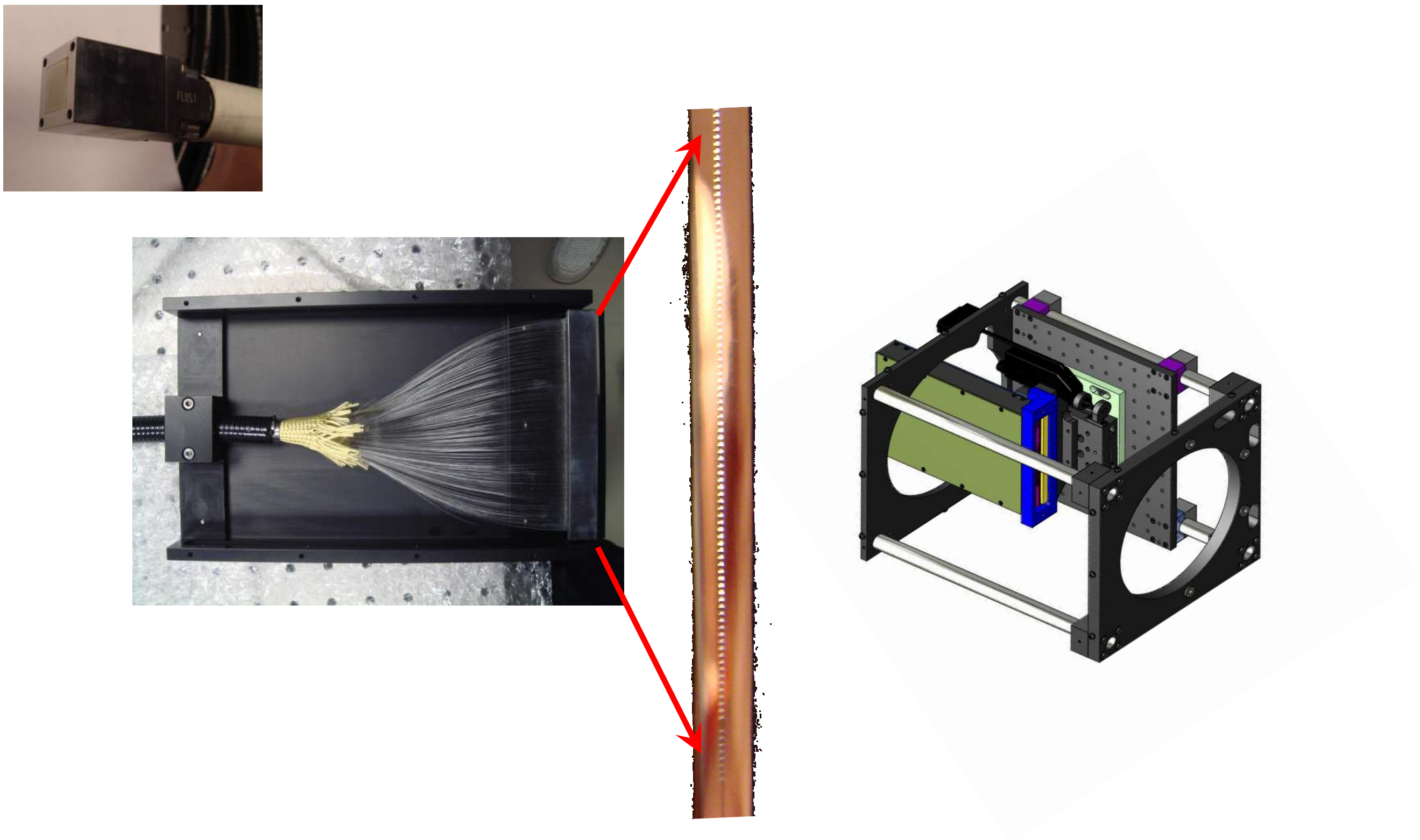

Figure 5: Fiber bundle input (top left), fiber-slit output (bottom) with expanded top view (arrows), MUSE spectrograph collimator interface with fiber-slit (CAD view, right)

The ERA2 bundle was fabricated by Leoni FiberTech in Berlin, Germany. It has a total of 400 multimode fibers with a core size of 110µm, forming a square matrix of 20×20 fibers with a pitch of 0.5mm and an overall footprint of 9.5mm × 9.5mm. As opposed to the VIRUS bundle where a large fill-factor is highly desirable, the ERA2 bundle is rather optimized to address a geometrically more extended sampling area at the expense of a small fill-factor – however, with the possibility to combine the bundle with a lens array and thus to achieve complete spatial coverage, or to maximize etendue for finite distance objects, respectively. A second bundle is available with 400 fibers of 80µm core diameter and a 20×20 matrix with 0.3mm pitch. The fiber bundles have an overall length of 10m. On the slit end, the MUSE spectrograph slit length of 130mm is filled with all 400 fibers, thus leading to a sparse packing and a pitch of 0.325mm. Unlike the initial VIRUS design, the fiber slit is not curved, but on a flat surface. Pictures of the fiber bundle, the fiber head, the fiber slit, and the MUSE spectrograph collimator interface are shown in Figs. 4 and 5, respectively. At the spectrograph input end, the fiber slit substrate is mounted on a protective cartridge that is compatible with the MUSE pinhole illumination unit and can be exchanged in a matter of a few minutes. The fiber head is made compatible with standard LINOS Mikrobank components in order to connect to lab experiments with a well-defined interface.

### 3.2 Lab setup of Raman imaging spectroscopy experiment

The first experiments to explore the capabilities of the ERA2 technology transfer demonstrator were prepared in collaboration with the Institut für Physikalische Chemie, Universität Jena. The objective of these experiments was to conduct a fast track *proof-of-concept* for multi-channel imaging Raman spectroscopy, using the available components of the ERASMUS-F test facility. Based on standard lab components, a setup was devised to feed the ERA2 fiber bundle with Raman-scattered light from a sample under investigation, as shown in Fig. 6.

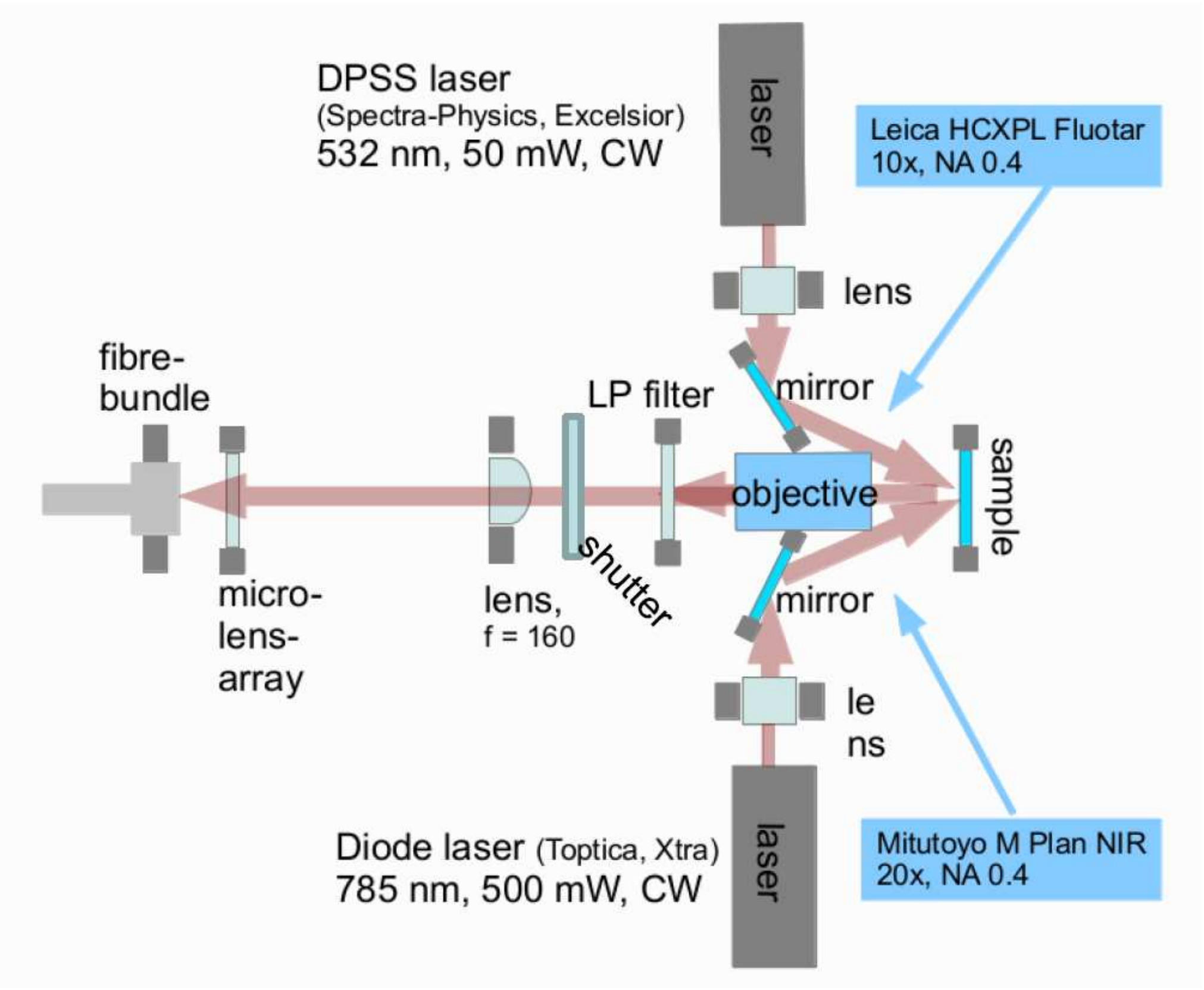

Figure 6: Laboratory setup for Raman imaging spectroscopy experiment

The setup incorporates basically an optical bench that optically couples a test sample to the ERA2 fiber bundle, using the combination of a microscope objective, a long-pass filter, a shutter, an imaging lens, a lens array, and finally the input head of the fiber bundle. There are two lasers devised to provide for coherent illumination with sufficient intensity at wavelengths of 532nm, 50mW, and of 785nm, 500mW, respectively. The corresponding illumination of the sample is controlled by a set of mirrors and lenses. A photograph of the whole setup is shown in Fig.7.

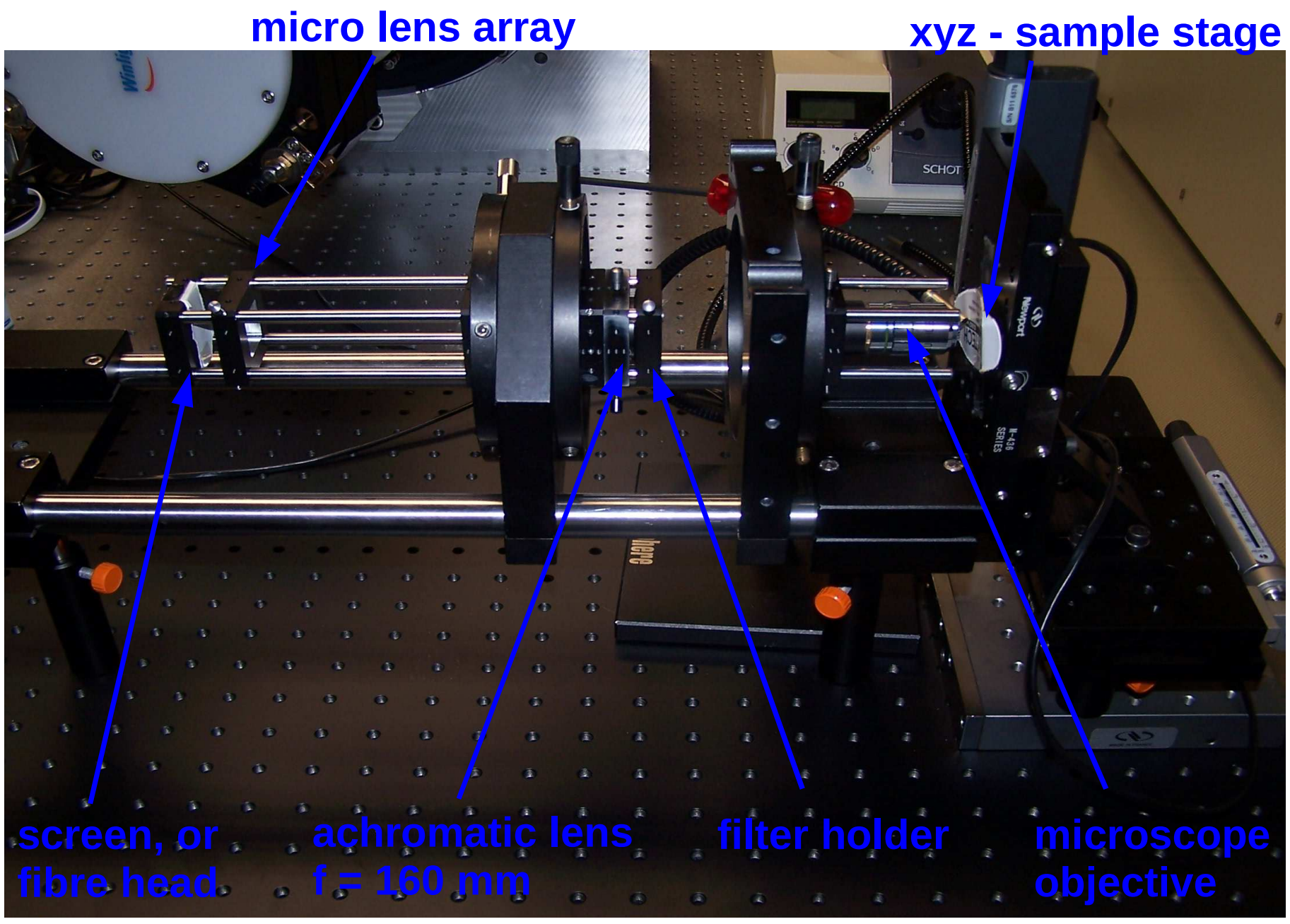

Figure 7: Photograph of lab setup for Raman imaging spectroscopy experiment

### 3.3 Experiments and first results

On May 3 and 4, 2012, first experiments were conducted with the setup as described above. A number of test samples were examined, including a Paracetamol tablet, minerals, and organic tissue. Here we report preliminary results for our reference case of the acetaminophen molecule contained in the Paracetamol tablet, which is expected from the literature to exhibit strong Raman scattering with a known spectrum. Fig. 8 shows a typical spectrum obtained from a 1 sec exposure of the sample using the 785nm laser. The figure is a screenshot from the p3d software tool that is commonly used for data reduction and visualization of integral field spectroscopy with fiber-based instruments. The program was originally developed for the PMAS instrument (Roth et al. 2005, Sandin et al. 2010), but subsequently generalized and adopted for other instruments as well (Sandin et al. 2011), see also Sandin et al. 2012 in these proceedings. p3d is a fully documented open source code that is distributed free of charge under GPLv3 from the Sourceforge platform http://p3d.sourceforge.net/. For the purpose of the ERA2 facility, p3d was adapted to accommodate the 400 fiber pseudo-slit with a position table that relates the location of a given fiber in the matrix of the input head of the fiber bundle to its location along the slit, i.e. the location of the corresponding spectrum on the CCD. With the aid of bias and calibration exposures using a setup where the target was replaced by an integrating sphere, fed either by arc lamps for wavelength calibration, or a tungsten filament source for flatfield calibration, the Raman CCD frames were fully reduced and visualized for analysis. The p3d graphical user interface as shown in Fig. 8 displays the row-stacked spectra as a false-colour plot with wavelength extending from left to right, and spectrum number according to location on the CCD counting from bottom to top (large frame in the upper part of the display). For a mouse-selected wavelength in this frame, the corresponding map at this wavelength is plotted in the frame to the upper right. For up to 10 such wavelength selections, the generated maps can be stored for inter-comparison in post-stamp format frames in the middle. Again for a mouse-selectable spatial element (“spaxel”) of the map display, the corresponding spectrum is displayed on the bottom as intensity vs. wavelength in analog-to-digital units of the detector system.

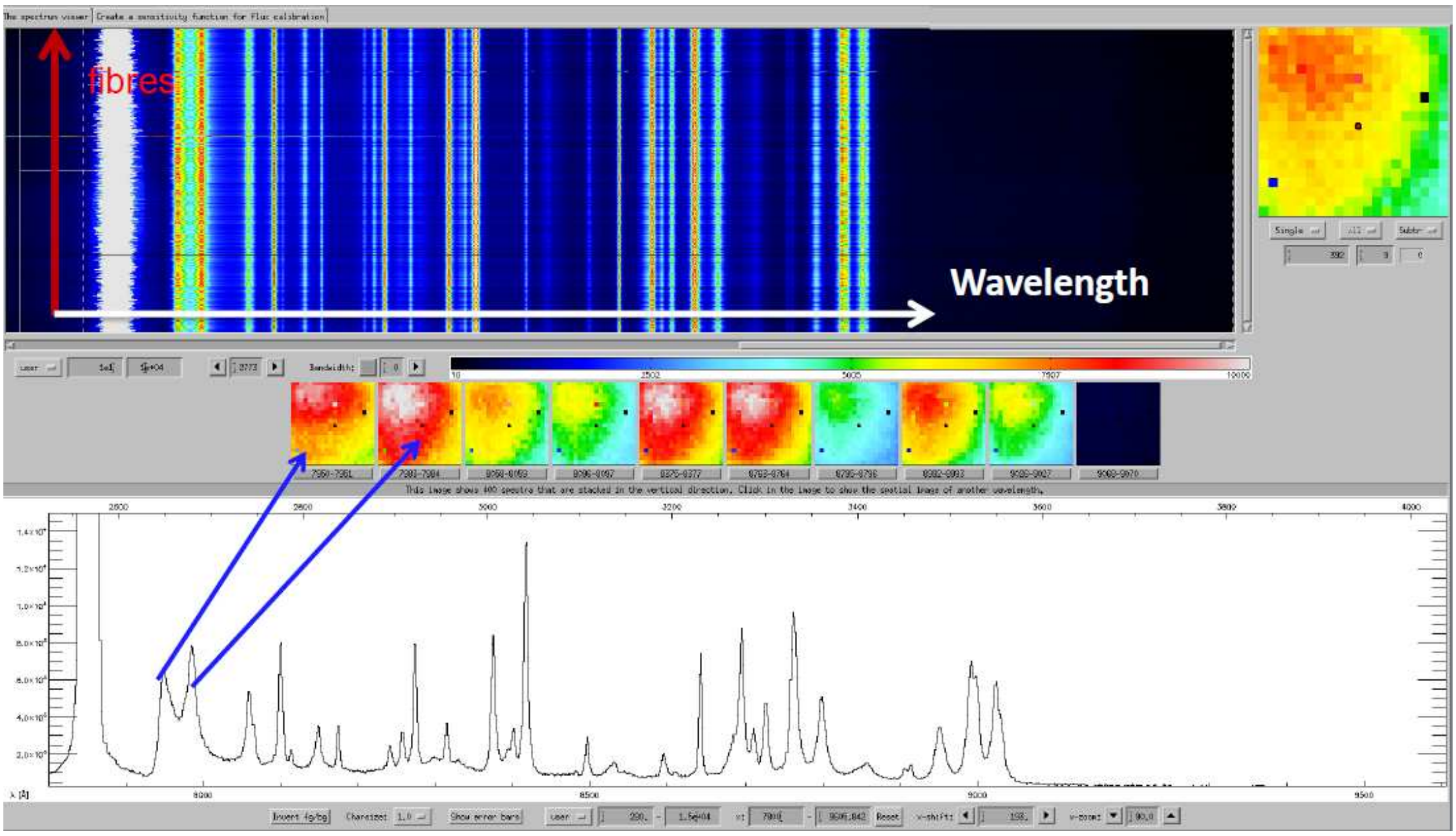

Figure 8: Graphical user interface of the p3d tool, showing the result of a Raman exposure (see text for explanation)

The spectra in Fig. 8 clearly show the presence of the residual laser line at 785nm that is still strongly visible despite the presence of the long-pass filter, and also a number of Raman emission features at different strengths with different spatial distributions as indicated in the maps for different wavelengths. The spectral images are decentered with respect to the geometry of the target: a clear gradient of intensity with a maximum near the upper left corner of the maps indicates the combined effects of a non-homogeneous laser illumination spot and the geometry of the tablet (note that dark spots and the less than perfect calibration of fiber-to-fiber transmission variations are indicative for the preliminary stage of data reduction). In order to validate the reality of the observed Raman features, we qualitatively compare the spectrum with the literature. Fig. 9 shows a comparison of our preliminary result with the spectrum of the acetaminophen molecule published by Pestaner et al. (1996), where a very good qualitative agreement is evident. Note e.g. the prominent feature at 799 $cm^{-1}$, that is known to be due to the stretch of the phenyl ring of the molecule, or the feature at 1649 $cm^{-1}$ owing to vibration of the CO bound, as illustrated in Fig. 9 (arrows).

## 4. SUMMARY AND CONCLUSIONS

In order to validate the potential of astronomical instrumentation for technology transfer, we have successfully performed a *proof-of-principle* experiment with a single MUSE spectrograph module, equipped with an imaging fiber bundle instead of a standard MUSE slicer. The bundle was coupled to an laboratory setup for Raman spectroscopy, immediately yielding meaningful results for a test case with acetaminophen that are compatible with the literature. We shall fully reduce and analyze our data obtained from the test campaign and publish the results in a forthcoming paper. As part of the joint activities of the *Leibniz Application Lab* of innoFSPEC with the ERASMUS-F project at AIP, we plan to further extend our promising first experiments towards a *proof-of-concept* demonstrator which is hoped to be of use in other disciplines.

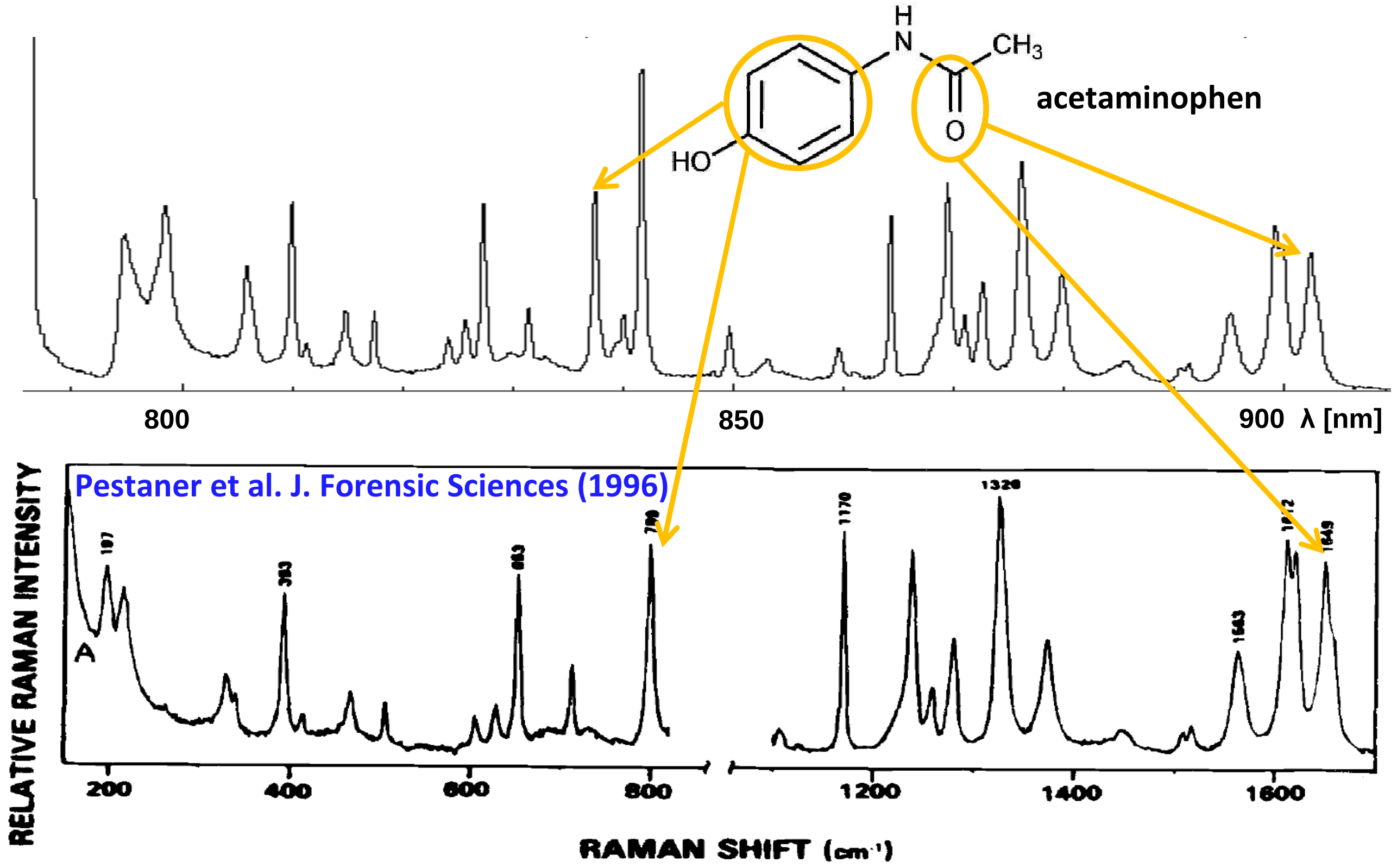

Figure 9: Raman spectrum of the acetaminophen molecule obtained from our ERA2 *proof-of-principle* experiment versus literature

## ACKNOWLEDGEMENTS

The Leibniz Application Lab was accomplished through financial support from the WtW funding programme of BMVBS under grant no. 03WWBB105. The authors also acknowledge support through funding from the SAW programme of the Leibniz Gemeinschaft in Germany, DLR grant no. 01SF1121, and from the ERASMUS-F project through funding from PT-DESY, grant no. 05A09BAA. We are especially indebted to Roland Bacon and the MUSE consortium for providing access to the MUSE detector vessel acceptance test facility at AIP for the purpose of the *proof-of-principle* experiment described in this article. The generous support of the European Southern Observatory with CCD detector equipment and the detector vessel is gratefully acknowledged.